\documentclass{nle}
\usepackage{graphicx}
\usepackage{placeins}
\usepackage{todonotes}

\usepackage[T1]{fontenc}
\usepackage[utf8]{inputenc}
\usepackage{tipa}

\title{Interpretability Analysis of Deep Models  for COVID-19 Detection}
\author[Silva et al. (2022)]{
\\
Daniel Peixoto Pinto da Silva\textsuperscript{1},
Edresson Casanova\textsuperscript{2},
Lucas Rafael Stefanel Gris\textsuperscript{3}, \\
Arnaldo Candido Junior\textsuperscript{4},
Marcelo Finger\textsuperscript{2},
Flaviane Svartman\textsuperscript{2},
Beatriz Raposo\textsuperscript{2}, \\
Marcus Vinícius Moreira Martins\textsuperscript{2},
Sandra Maria Aluísio\textsuperscript{2},
Larissa Cristina Berti\textsuperscript{4}, \\
João Paulo Teixeira\textsuperscript{5} \\ \\
\textsuperscript{1} Federal University of Technology - Paraná \\
\textsuperscript{2} University of São Paulo \\
\textsuperscript{3} Federal University of Goias \\
\textsuperscript{4} São Paulo State University \\
\textsuperscript{5} CeDRI - Instituto Politécnico de Bragança
}

\received{24 Nov 2022}

\pagerange{\pageref{firstpage}--\pageref{lastpage}}
\pubyear{1998}

\begin{document}

\label{firstpage}
\maketitle

\begin{abstract}
During the outbreak of COVID-19 pandemic, several research areas joined efforts to mitigate the damages caused by SARS-CoV-2. In this paper we present an interpretability analysis of a convolutional neural network based model for COVID-19 detection in audios. We investigate which features are important for model decision process, investigating spectrograms, F0, F0 standard deviation, sex and age. Following, we analyse model decisions by generating heat maps for the trained models to capture their attention during the decision process. Focusing on a explainable Inteligence Artificial approach, we show that studied models can taken unbiased decisions even in the presence of spurious data in the training set, given the adequate preprocessing steps. Our best model has 94.44\% of accuracy in detection, with results indicating that models favors spectrograms for the decision process, particularly, high energy areas in the spectrogram related to prosodic domains, while F0 also leads to efficient COVID-19 detection.
\end{abstract}

\section{Introduction}
In December 2019, a new coronavirus species was found, named SARS-CoV-2, responsible for the COVID-19 (Coronavirus disease 2019). This variant affects humans and rapidly became a worldwide worry, as it reached the status of Pandemic according to WHO (Word Health Organization) \cite{whospeech}. This virus evolved and became more contagious and lethal in a short period of time.

% https://www.ncbi.nlm.nih.gov/pmc/articles/PMC7569573/
% https://www.bbc.com/news/world-51839944

tackle the pandemic crisis. In particular, researchers from Artificial Intelligence and related areas searched for methods to simplify COVID-19 detection. These methods uses a variety of sources, such as exams \cite{brinati2020detection}, symptoms \cite{zoabi2021machine}, x-ray images \cite{ozturk2020automated}, among others \cite{acar2021improving}. An potential source to COVID-19 detection is the use of audio. Several projects have collected audios from patients in the form of speech or cough \cite{brown2020aexploring,brown2020bexploring,casanova-etal-2021-deep}. Such models may be useful to optimize patient screening. However, existing approaches have limitations in the capture method. For example, environmental noises may be present during audio capture. In this context, it is easy for a model to overfit on such noises. 

For example, the dataset presented in SPIRA Project \cite{casanova-etal-2021-deep} was built collecting positive audio samples (read speech) from COVID-19 patients in hospitals, while asymptomatic examples\footnote{This examples were labelled as control group, although no additional testing for COVID-19 were performed on those subjects.} were donated via Web. Training a model on this dataset should incorporate precautions towards learning biases due to collection environment differences, as patients audios may contain hospital noises while the control group may contain other environmental noises.

In this work, we train and analyze Convolutional Neural Networks (CNNs) for COVID-19 detection from audios using the dataset from \cite{casanova-etal-2021-deep}. We analyze which factors are important for the model decisions using several criteria, namely spectrograms, fundamental frequency (F0), fundamental frequency standard deviation (F0-STD), speaker age and sex. We apply the algorithm Grad-CAM (Gradient-weight Class Activation Mapping) \cite{gradVisu} to generate heat maps to investigate which pieces of information are more relevant in the decisions. As the dataset used in this work contains audios from different collection environments (hospital and domestic), learning biases can occur towards hospital noises. To mitigate this problem, we inject hospital noises in domestic audios following the proposal of \cite{casanova-etal-2021-deep}. We also use other methods for improving the model performance, including transfer learning and data augmentation. Finally, we can literally hear the areas in the audio that the model value the most in the its decision process. In order to do that, heat-maps obtained by Grad-CAM are multiplied by the original log-mel spectrograms and the result is synthesized.

This work presents three main contributions:

\begin{enumerate}
    \item We present an analysis detailing which features are important for neural models to detect or discard COVID-19 in patient and control audios.
    \item We present an interpretation of the decisions made by deep models using heat maps and audio synthesis, following an approach of explainable Artificial Intelligence.
    \item We show that, given suitable preprocessing steps, deep models can taken unbiased decisions even when the audio capture process contains spurious data such as hospital wards noises, given difficulties  in audio capture due to the circumstances of COVID-19 pandemic.
\end{enumerate}

\section{Related Work}

In related work, COVID-19 is detected using different approaches, considering the input features used as a base for the classification. From the perspective of feature analysis, this inputs can be roughly grouped in two categories, white-box or black-box, according to their easiness of interpretation.

An example of approach using mostly white-box features is the work of \cite{bartl2020voice}. The authors use 88 features extracted from audios containing vowels in order to measure how their differ from COVID-19 patients to the control group. The authors found that F0-STD commonly varies between these two groups. In this work, we also use of F0, F0-STD, but we include sex and age as input for our deep models to detect COVID-19. The work of \cite{Hberti2021changes} previous study identified that sex and age influence F0 and F0-STD in COVID-19 patients, and that women and elderly subjects presents more differences in these two parameters, as their voice becomes higher-pitched and less stable.

Regarding black-box features, the work of \cite{schuller2021interspeech} proposes a challenge for COVID-19 detection from both speech and cough audios using the Cambridge COVID-19 Sound database \cite{brown2020aexploring,brown2020bexploring}. The authors perform baseline experiments and indicates thousands of features that can be used to general audio processing and, in particular, COVID-19 detection in audios.

MFCCs (Mel\-Frequency Cepstral Coeficients) \cite{zheng2001comparison} proved to be an useful method COVID-19 detection while consuming few computational resources \cite{casanova-etal-2021-deep}. More robust approaches use spectrograms, transfer learning and data augmentation for the task \cite{casanova2021transfer}. Based on this results, we also investigate the use of spectrogram for COVID-19 detection, as well as transfer learning and data-augmentation.

Related work either uses white-box features (ex.: sex and age) to understand COVID-19 effect on audios or black-box features (ex.: spectrograms) for better classification results. In this work, we propose an interpretability analysis of the decisions made by  deep models found in the literature, in pursuance of better understanding of their results. In order to do this, we propose the use of the algorithm Grad-CAM \cite{gradVisu}, analysing its heat maps and synthesising audios based on this heat maps. 

\section{Method}

Our data are obtained from the SPIRA dataset. We use a balanced version of the dataset containing spoken utterances from 432 speakers including patients and control group members. Audios from patients are collected when their blood oxygenation level is bellow 92\%. The dataset was divided into training (292), validation (32) and test (108). The dataset has recordings of patients and control group speaking the utterance ``o amor ao próximo ajuda a enfrentar o coronavirus com a força que a gente precisa'' (``love of neighbour helps to face coranavirus with the strength we need'').

In this work, we tested transfer learning from pretrained models (Section \ref{sec:transfer}), data augmentation techniques (Section \ref{sec:augmentation}), audio splitting based on windowing (Section \ref{sec:windowing}), preprocessing during training steps (Section \ref{sec:dyn}), which we called dynamic preprocessing. This approach lead to six experiments (sections \ref{sec:exp1} and \ref{sec:exp2}).

\subsection{Transfer Learning with Pretrained Audio Neural Networks}
\label{sec:transfer}

Pretrained Audio Neural Networks (PANNs) has proven adequate for transfer learning as it can be used in a wide variety of tasks \cite{kong2020panns}. The model we used is mel spectrogram based and was trained over the AudioSet dataset which have approximately 1.9 million audios, totalizing 527 classes and over 5,000 hours. Although the authors explored some architectures, in this work we only applied the CNN14 since it is one of the simplest PANNs. Also, this model is similar to SpiraNet \cite{casanova-etal-2021-deep}, and can benefit from the same preprocessing techniques.

\subsection{Data Augmentation}
\label{sec:augmentation}
Similar to the work of \cite{casanova2021transfer}, three data augmentation techniques were applied: noise insertion, Mix-up and SpecAugment.

Noise insertion was performed because the SPIRA dataset has different recording environments for patients and control group audios. \cite{casanova-etal-2021-deep} showed that models trained over this dataset can overfit if the data are not prepared adequately. A model trained over the original data can easily bias, separating control and patients by the presence of hospital ward noises. In order to prevent that, \cite{casanova-etal-2021-deep} proposes the use of noise injection in all audios. In this work, we follow the same approach. For some of our proposed experiments, 4 noise recordings are inserted for control and 3 for patients, while other experiments use 3 audio recordings for each class. %Noise recordings are constantly sampled and injected in audios for each step of the training process (dynamic noise injection), in order to prevent overfit.

Mix-up  consists of combining two random instances ($x_i$ and $x_j$) from the training set and their respective classes ($y_i$ and $y_j$) to generate a new instance ($\tilde{x}, \tilde{y}$) \cite{zhang2017mixup}. This instance is generated using equations \ref{eq:mixupX} and \ref{eq:mixupY}, where $\lambda \in [0,1]$ is generated from a Beta distribution. This data augmentation technique can be used on a wide variety of tasks.

    \begin{equation}
        \label{eq:mixupX}
         \tilde{x} = \lambda x_i + (1 - \lambda)x_j
    \end{equation}
    \begin{equation}
        \label{eq:mixupY}
         \tilde{y} = \lambda y_i + (1 - \lambda)y_j
    \end{equation}
    
The method differs from other data augmentation techniques such as rotation, cropping and horizontal flipping that are more common techniques in image processing. Also this technique can be used in several tasks, including audio processing \cite{xu2018mixup}.

SpecAugment \cite{park2019specaugment} is focused on data augmentation on spectrograms. It was designed for automatic voice recognition. It performs the augmentation on the spectrogram first by distorting it in the time dimension, and then masking parts of frequency channels and masking blocks in time. The frequency mask is made over $f$ consecutive Mel channels $[f_0,f_0+f)$, where $f$ is chosen from a uniform distribution from 0 to $F$, and $F$ represents the maximum number of masked mel channels (8 in our experiments). The parameter $f_0$ is obtained from $[0,v-f)$ where $v$ is the number of mel frequency channels. The temporal mask is performed over $t$ time slots $[t_0,t_0 + t)$, the calculations for this mask are analogous for the previous mask.

\subsection{Windowing}
\label{sec:windowing}

Each original audio is divided into smaller four-second audios. To do this, we use a four-second window with 1 second hop. For example, an audio with 5 seconds is split in two: one from seconds 0-4, and other for seconds 1-5. This is performed to make the audio length uniform. Since patient audios tend to be longer, models can overfit on audio length if this normalization is not done.

The window cover repeated fragments of the original audios in order to include many fragments of the original spoken sentence as possible in each generated audio. It is important to note that windowing is done separately for training and test. In the test set, voting over the windowed audios is then used to decide the class of an original audio. The voting is the sum of the predicted probabilities for each class. Windowing also works as a simple data augmentation technique, in addition to the approaches presented in Section
\ref{sec:augmentation}.

\subsection{Dynamic Preprocessing}
\label{sec:dyn}

The audios are preprocessed again for each training step, ensuring a richer variety in augmented data. To maintain our model consistent, the same applies for validation and test. The following operations are carried out:

\begin{enumerate} 
  \item Noise Injection; 
  \item Windowing; 
  \item Spectrogram extraction; 
  \item Spec-augment application (only for training); 
  \item Mix-up application (only for training);
  \item Training step / test step.
\end{enumerate}

Operations 4 and 5 are only applied to PANN based experiments, and only in training, while the others operations are common for all experiments. Regarding operation 3 we used different parameters for spectrogram extraction in our experiments. Table \ref{tab:settings} presents the two settings used across experiments presented in section \ref{sec:exp1} and \ref{sec:exp2}. Two parameters were common for all experiments based on spectrograms: the number of Fast Fourier Transform \cite{brigham1967fast} components (1,200) and the spectrogram format (log-mel). The other parameters changed according Table \ref{tab:settings}.

%The log-mel spectrogram was extracted using Fast Fourier Transform (FFT) \cite{brigham1967fast}, with Hamming window 1024, hop length 320 and 1024 FFT components

% o exp de data aug usa mel normal
% other  hop 160 win 400 nfft 1200 numfreq 601 num mel 80
% data aug hop 160 win 400 nfft 1200 numfreq 601 num mel 40
% pann hop 320 win 1024 nfft 1200 numfreq 513 num mel 64

\begin{table}[htb]
\caption{Settings used in the Experiments}
\begin{tabular}{lrrrr}
\hline
\textbf{Set} & \textbf{Hop Size} & \textbf{Num Freq} & \textbf{Num Mel} & \textbf{Win Len} \\ \hline
1                                & 160                                   & 601                                   & 80                                   & 400                                  \\
2                                & 320                                   & 513                                   & 64                                   & 1024   \\ \hline                             
\end{tabular}
\label{tab:settings}
\end{table}

\subsection{Experiments Over Inputs}
\label{sec:exp1}

\begin{figure}[htb]
\centering
\includegraphics[width=0.8\textwidth]{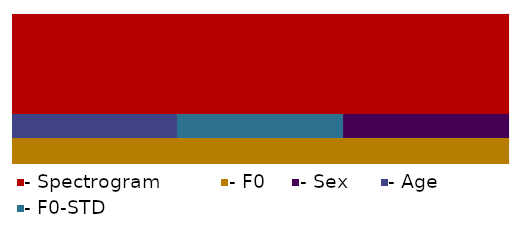}
\caption{Input representation}
\label{fig:rep}
\end{figure}

The first three experiments investigate the role of different information (spectrogram, F0, F0-STD, age and sex) in the model decisions process. Spectrograms are matrices, while F0 is a vector and the remaining data are scalars. We decided to convert all these data into matrices, as our purpose is to analyse visually the importance of each input using Grad-CAM. In order to do that, we propose the representation showed in Figure \ref{fig:rep}. The input in its full form has $401\times120$ pixels, where the spectrogram occupies the top region ($401\times80$). Next age, F0-STD and sex consumes 20 lines, while consuming either 133 (age and sex) or 135 columns (F0-STD). Age pixels have all the same shade of grey, considering this input is an scalar. The same applies to F0-STD. Sex is simpler, because is binary scalar, assuming only zero (male) or one (female). Finally, F0 is represented in a ``bar code'' style, where each value in the original vector is repeated through a whole column in the generated matrix.

Using the scheme presented in Figure \ref{fig:rep}, three first proposed experiments are:

\begin{itemize}
\item Experiment 1: based on spectrogram only ($80\times401$ pixels)

\item Experiment 2: uses F0, F0-STD, age and sex ($40\times401$ pixels).

\item Experiment 3: uses all input data present in Figure \ref{fig:rep} are used ($120\times401$ pixels).

\end{itemize}

All three experiments are based on the SpiraNet model and their configurations are from Set 1 of Table \ref{tab:settings}. 

\subsection{Experiments over Training Process }
\label{sec:exp2}

Three experiments related to changes in the training process, pre-training and post-processing were performed:

\begin{itemize}
\item Experiment 4: focuses on pre-training, exploring the use of transfer learning using a PANN model (CNN-14). It was configured using Set 1 from Table \ref{tab:settings}.

\item Experiment 5: explores the training itself and uses dynamic data augmentation based on SpecAugment and Mix-up. It was configured using Set 2 from Table \ref{tab:settings}.

\item Experiment 6: analyzes post-processing of the results, generating new audios. This is done in two step. First, the heat-map generated by Grad-CAM and the log-mel spectrogram are combined using the Hadamard product. Second, the result and the phase of the original spectrogram is used to generate new audios highlighting the moments in time and the frequencies that the model considered the most in a given decision. We refer to the combination of original log-mel spectrograms with heat map as modified spectrograms. This experiment was based on our best model (PANN CNN14 from experiment 4).
\end{itemize}

\section{Results and Discussion}

\subsection{Experiment Results}
\label{sec:exp_results}

Table \ref{tab:results} presents the results of experiments 1 to 5. Experiment accuracies varied from 65.75\% to 94.44\%. From results of experiment 1, we observe that spectrograms are discriminative. Likewise, experiment 2 showed that F0, F0-STD, sex and age also contain discriminative information. However, spectrograms seems to carry more useful information since its accuracy in experiment 1 is more than 10\% superior than in experiment 2. Experiment 3 suggests that features extracted from inputs in experiments 1 and 2 are mostly equivalent, despite a slight increase in accuracy (~2\%) compared to experiment 1.

\begin{table}[htb]
\caption{Results from Experiments 1 to 5}
\begin{tabular}{llllll}
\hline
\textbf{}     & \textbf{True}      & \textbf{True}      & \textbf{False}     & \textbf{False}     & \textbf{}         \\
\textbf{Exp.} & \textbf{Positives} & \textbf{Negatives} & \textbf{Positives} & \textbf{Negatives} & \textbf{Acc (\%)} \\ \hline
1             & 37                 & 49                 & 5                  & 17                 & 79.63             \\
2             & 36                 & 38                 & 16                 & 18                 & 68.52             \\
3             & 44                 & 44                 & 10                 & 10                 & 81.48             \\
4             & 51                 & 51                 & 3                  & 3                  & 94.44             \\
5             & 49                 & 22                 & 32                 & 5                  & 65.74             \\ \hline
\end{tabular}
\label{tab:results}
\end{table}

Experiment 4 resulted in the highest accuracy (94.44\%). This result indicates that transfer learning has a major impact in learning features from patients and control group, surpassing the results in the original work of \cite{casanova-etal-2021-deep}. It also indicates that CNN14 would be better suitable than SpiraNet. Experiment 5 indicated that data augmentation techniques (SpecAugment and Mix-up) did not improved the learning process, as it was worse than experiments 1 and 2. Experiment 6 is presented separeted since it is based on audio resynthesis and human analysis (sections \ref{sec:preliminary} and \ref{sec:detailed}).

Regarding errors, most experiments resulted in a balance of false positives and false negatives. Experiment 1 was one exception, presenting more false negatives. Maybe this experiment was more susceptible than others to cases of silent hypoxia, in which a patient have low blood oxygenation, but does not present strong symptoms. The other exception was experiment 5, in which there are much more false positives (32) than false negatives (5). An hypothesis to this phenomenon is the spec augment force the model to give less importance to pauses, which is an important feature to detected respiratory insufficiency. This may occur because the method inject artificial pauses in training data.

\subsection{Hypotheses for Model Decision Process}

During a group preliminary experiments focusing on spectrogram analysis, we noted that model training can lead to different learning aspects. This result is expected because artificial neural networks are high-variance, low-bias classifiers with randomized parameter initialization. We also noted that transfer learning tends to reduce the model variance. In our analysis, we formulated hypotheses to explain the obtained variance and to understand the data aspects that may play a role in model learning:

%1. Modelo usa pausas para a classe positiva
%2. Modelo usa vogais baixas para a classe negativa
%2.1 Modelo foca nelas porque eles tem mais energia, especialmente F1 e F2
%2.2 Elas são pronunciadas com mais energia por causa dos níveis linguísticos superiores (morfossintático e semântico-pragmático influenciam a prosódia)

\begin{itemize}

\item $H_1$: pauses are important clues to detect COVID-19 since the patients tend to make more pauses for breathing than the control group.
%está estranho este fraseamento. Vou refrasear sem a palavra chave antes.

\item $H_2$: as the air starts decreasing in the lungs, the speaker may begin to loose his breath or the signal energy may begin to decrease. Thus energy over time can be an important clue.

\item $H_3$:  an interplay between syntax and prosody is expected to emerge as a boundary marked by formant vowel high energy, i.e. phonetically.

%hipóteses excluídas:

%\item Formant region ($H_3$): as formants are high energy regions in the signal, they may indicate lung capacity issues.

%\item Prosodical features ($H_4$): tonic syllables may be more demanding to the vocal apparatus and, specially, to the lungs.
%Prosodical features may be strategic to COVID-19 detection considireing not only tonic syllables, but pre-tonic and post-tonic as well. 

%\item Other languages levels ($H_5$): there is an interplay between syntax and semantic-pragmatic context in prosody. Therefore, these relationships may play a role on relevant features for classification.

\end{itemize}

% exemplo: gradcam para gato e cachorro (cada vez olha um traço diferente do animal)

% o uso de técnicas como transfer learnig e data-augmentation ajudam a reduzir um pouco essa variância

% nossas redes aprenderam coisas diferentes
% neste trabalho, vamos "interpretar" a rede mais interessante

% com base nessas informações, escolhemos o experimento {1,4}

\subsection{Preliminary Analysis: Heat Map Case Study}
\label{sec:preliminary}

\begin{figure*}[!htb]
\centering
\includegraphics[width=0.9\textwidth]{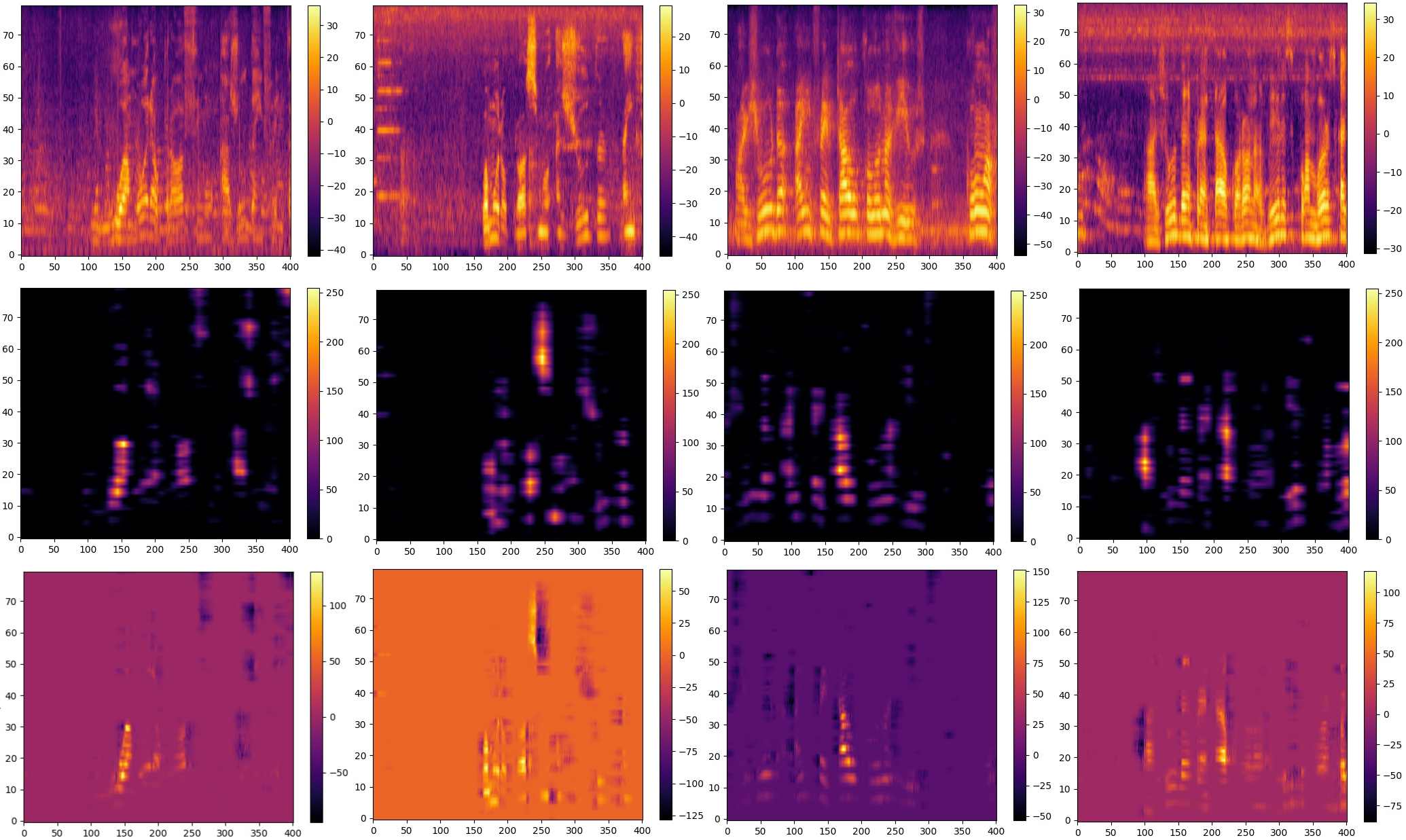} \caption{Results from Experiment 1 (spectrogram only) including original spectrograms (top), heat maps (middle) and modified spectrograms (bottom) for two control group members (left) and two patients (right).}
\label{fig:exp1}
% \arnaldo{pegar figuras do tcc novamente transposto com linha 1 espectrogramas; linha 2 mapas; linha 3 produtos. obs: figura acima é só rascunho}
\end{figure*}

\begin{figure*}[!htb]
\centering
\includegraphics[width=0.9\textwidth]{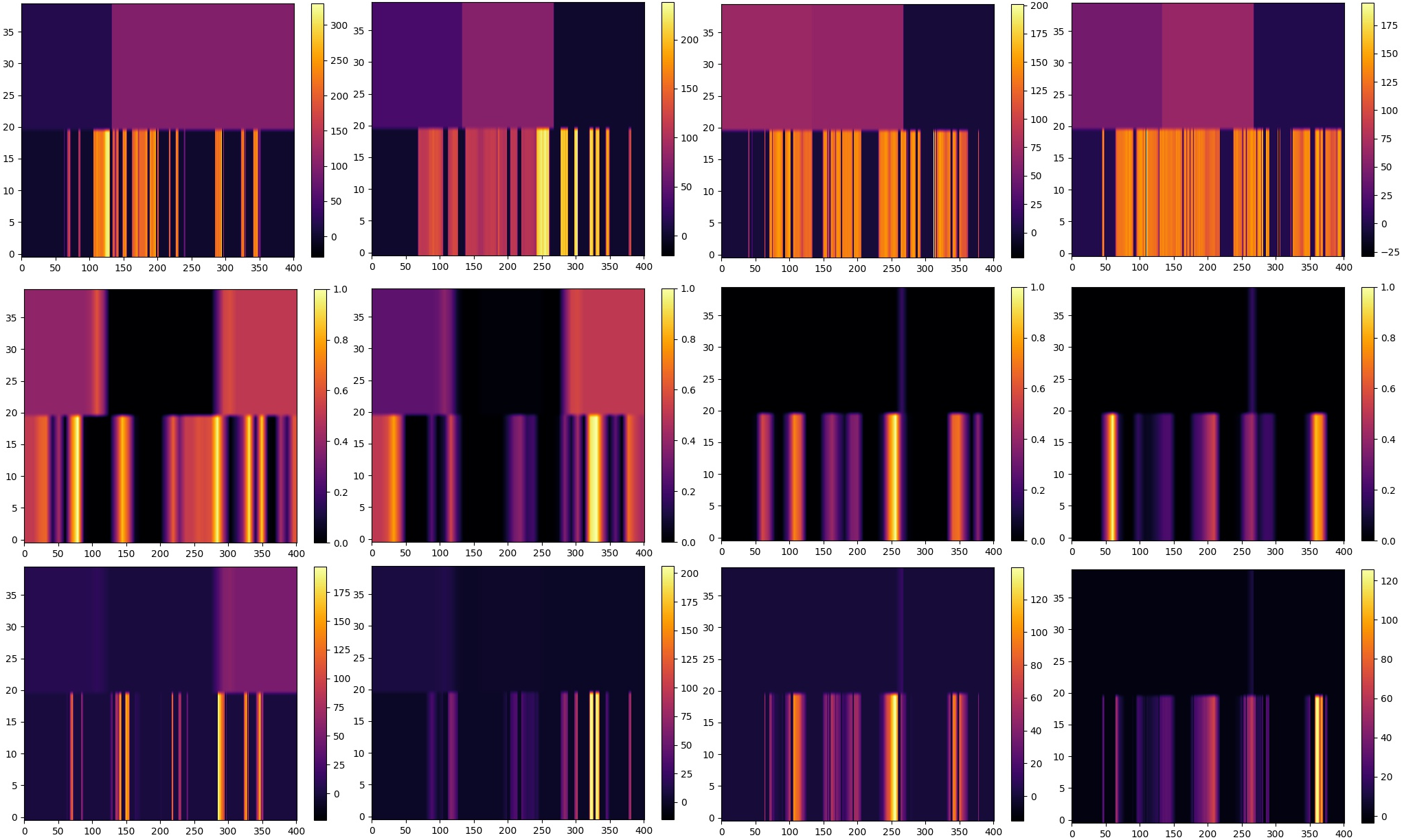} 
\caption{Results from Experiment 2 (all inputs except spectrograms) including original images (top), heat maps (middle) and modified images (bottom) for two control group members (left) and two patients (right).}
\label{fig:exp2}
\end{figure*}

\begin{figure*}[!htb]
\centering
\includegraphics[width=0.9\textwidth]{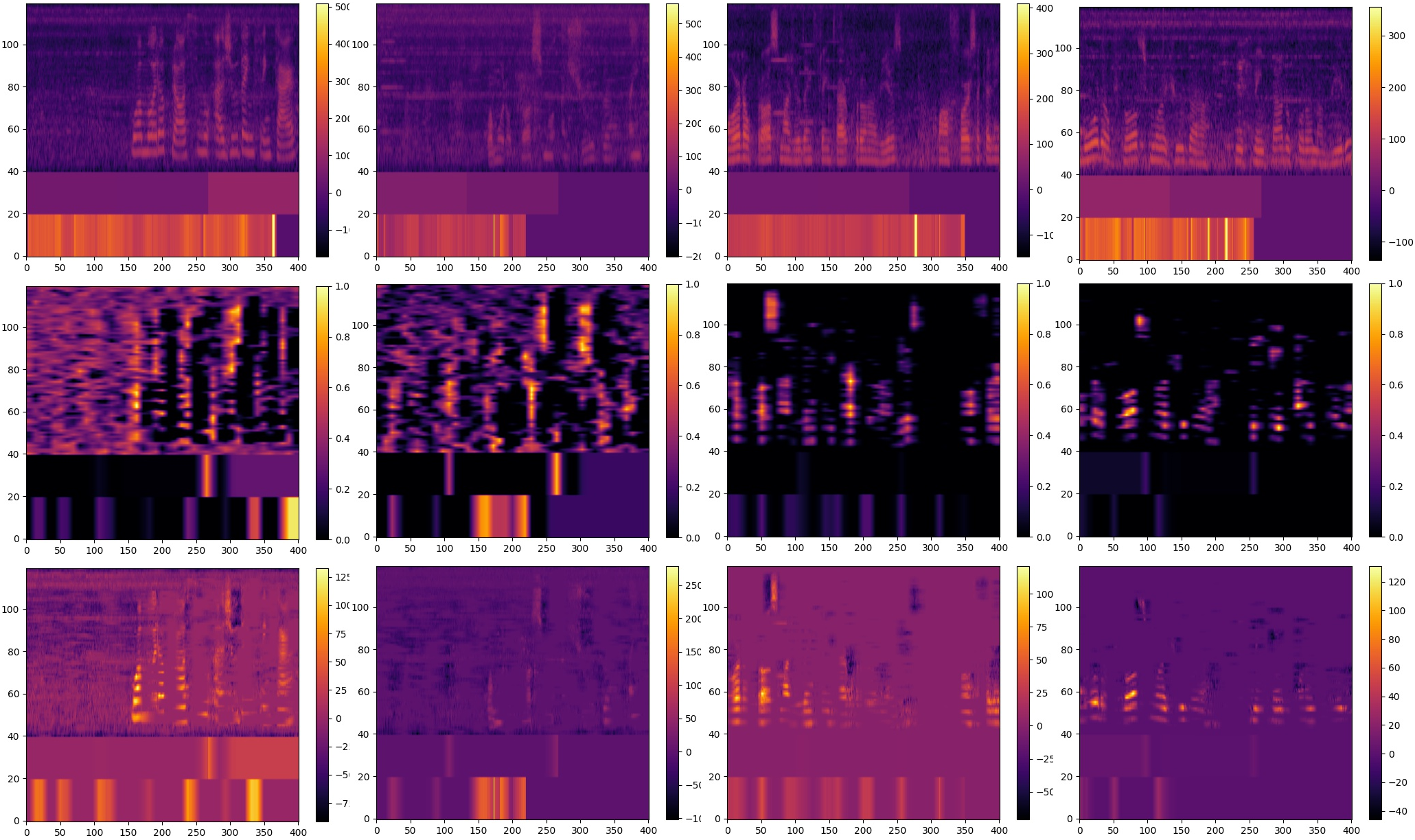} 
\caption{Results from Experiment 3 (all inputs) including original images (top), heat maps (middle) and modified images (bottom) for two control group members (left) and two patients (right).}
\label{fig:exp3}
\end{figure*}

Experiment 6 is different from the others as it involves human evaluation. We made two analyzis: a case study (this section); and a detailed analysis (Section \ref{sec:detailed}). These analyzes were performed using Grad-CAM, as we generated heat maps for experiments based on inputs (Section \ref{sec:exp1}). Figure \ref{fig:exp1} to  \ref{fig:exp3}  presents results for experiments 1 to 3, respectively.

Our first experiment was based on spectrograms only. The visual results for two patients and two control group members, considering original spectrograms, obtained heat maps and modified spectrograms are presented in in Figure \ref{fig:exp1}. We observe activity (attention) in high energy regions over the input. These results suggest that energy levels and audio formants ($H_2$ and $H_3$) may play a relevant role in COVID-19 detection.

Previously, in Section \ref{sec:exp_results}, the results indicated F0, F0-STD, age and sex as distinctive features for COVID-19 detection. Figure \ref{fig:exp2} presents Experiment 2 visual representations for two patients and two control group members. It can be noted that F0 plays a major role in this model detection process. Specially in regions associated with transitions from  voiced phonemes to pauses ($H_1$) or to voiceless phonemes. The same applies to transitions from pauses or voiceless phonemes to voiced phonemes. Also, sex and age seem to be play in control classification, although not as noticeable as F0. F0-STD, on other hand, seems to be simply disregarded by the model.

Figure \ref{fig:exp3} presents the visual representations generated using all information available (spectrograms, F0, F0-STD, sex and age). Heat-maps suggest that spectrograms, F0 and sex are useful for patient classification, while control group detections are based on only on spectrograms and F0. This indicates spectrograms and F0 may contain complementary information given the slight superior accuracy obtained from Experiment 3 when compared with experiments 1 and 2.

\subsection{Detailed Analysis: Phonetical Investigation}
\label{sec:detailed}

The analysis presented in this section consists on four main inputs:

\begin{itemize}
\item Regular spectrograms in Hertz obtained from original audios. They were generated using the software PRAAT v6.1.09.
\item Regular spectrograms in Hertz from audios resynthesized from our modified spectrograms. These spectrograms combines speech with heat-maps.
\item Original and modified mel-spectrograms as presented in Section \ref{sec:preliminary}
\item Resynthesized audios from the modified spectrograms. These audios allow us to \textit{hear} where the model pays attention, while spectrograms allow us to \textit{see} where the model focuses. The audios are publicly available\footnote{https://drive.google.com/drive/folders/1aQEq82iUpnAmrQzQ52458GORv8PEK3nr?usp=share\_link}.
\end{itemize}

As our windowing approach consisted in generating four-second windows with one-second hop, this analysis considered only the central audio fragment in the window. In total, the central fragment of 73 audios were inspected, containing 30 correct predictions for each class and 13 errors (7 patients and 6 controls).

It was observed that the decision process usually falls on two aspects of the speech sound signal: the continuity of the signal versus its interruption. Thus, the model seems to pay attention to an alternation of continuity of speech sounds as opposed to their discontinuity, which in terms of intonational analysis are pauses inserted by speakers. This is in line with the finding by \cite{fernandes2022temporal}, in which patients' pauses are significantly longer than those of control subjects and, being in greater number, are inserted in more places throughout the utterance.

Considering short-term parameters, such as the most salient vowels for the model, it was observed that the vowels /a/ from ``ajuda a'' (``helps to'') and ``enfrentar'' (``to face''); /\textopeno/ from ``próximo'' (``neighbor''); /o/ from ``força'' (``strength''); /oN/ from ``com'' (``with'') are those reproduced with more intensity in the modified spectrogram. This corresponds to what occurs in the original audio spectrogram, and besides being expected, it is reasonable, since these vowels occupy prominent places in the utterance, or are intrinsically more intense, such as the low vowels /a/ and /\textopeno/. On the other hand, the mid-high, oral /o/ and nasal /oN/ vowels do not have the same sound amplitude as the low ones, but occupy a place in the utterance that prosodically highlights them.

Therefore, on one hand, we have the phonetic features of vowels and, on other hand, the prosodic feature interacting with morpho-syntactic and semantic-pragmatic levels. The interaction discussed here explains the emphasis on the verb (there is usually a peak of F0 in the verbal item in a statement). In semantic terms, emphasis is given to ``com a força''  (``with the strength''). The initial expression of the adverbial phrase ``com a força que a gente precisa'' (``with the strength we need'') is often phrased as an intonational phrase in our data. The intonational phrase initial position is prominent in Portuguese \cite{frota2014prosody,tenani2003dominios}. In pragmatic terms, this adverbial phrase is new information that modalizes the meaning of the verb “to face” (it is necessary to face the virus with strength).

Taking into account the phonetic analysis in interaction with other linguistic levels, the following can be said about the model preferences to distinguish patients from controls:

\begin{enumerate}
    
\item Average patient pauses (approximately 400 ms) represent large interruptions of formant frequencies tracks.

\item Vowels intrinsically more intense are important clues. Such vowels had the first (F1) and the second (F2) formants highlighted by the model in a 550 to 1300 Hertz range. This occurred for both groups, and for patients, such highlight can come before or after pause.

\item The interaction between non-low vowels (with F1 and F2 ranging from 350 to 1000 Hz), the morpho-syntactic context, the prosodic domain in which they are produced and its semantic-pragmatic role indicate that these segments are more prominent in the utterance, which seems to be important for the model.

\item The interaction between non-low vowels and the initial position in the intonational phrase (``com a força que a gente precisa'' -- ``with the strength we need'') results in prosodic emphasis of this unit (``com a força'' -- ``with the strength''), which draws the model attention.

\end{enumerate}

Finally, for the correct predictions, regarding the interplay between the speech sound signal and the prosodic behavior that emphasizes it, we propose the following explanatory hypothesis: the model sees the signal as continuous and with emphasis on some vowel formants in one of the groups; in another group, it sees important interruptions in a similar speech signal (the same sentence uttered by patients and controls), which leads to successfully distinguishing two different speech groups, since patients, with respiratory difficulties, are unable to produce fluent speech, and usually speaks the linguistic utterance with many pauses.

%1. Modelo usa pausas para a classe positiva
%2. Modelo usa vogais baixas para a classe negativa
%2.1 Modelo foca nelas porque eles tem mais energia, especialmente F1 e F2
%2.2 Elas são pronunciadas com mais energia por causa dos níveis linguísticos superiores (morfossintático e semântico-pragmático influenciam a prosódia)

%\FloatBarrier

\section{Conclusions}

This work presented an method for interpretability analysis of audio classification for  COVID-19 detection based on convolutional neural networks. Our work focus on explainable Artificial Intelligence. We investigated the importance of different features in the training process and generated heat maps to understand model reasoning for its predictions.

Regarding the input data, our results showed that spectrograms are a suitable representation for COVID-19 detection. F0 seems to be almost as efficient as spectrograms, and the combination of the two inputs lead to an small increase the model performance. Our best model achieved 94.44\% of accuracy. The heat maps and resynthesized audios support the claim that our models are not using environmental noises into account in order to make decisions, rather focusing on important factors on the signal, such as energy level and formants. From a linguistic perspective, prosodic boundaries are important cues to distinguish the groups of speakers with the most important features being pauses and formant energy in the beginning of the intonational phrases.

In future works, we plan to investigate other audio related features, such as autocorrelation, jitter and shimmer. We also intend to investigate the sentence start, when a speaker starts to produce a sentence, he/she has more air in the lungs, which starts to decrease as the person speaks. Some models may focus more on the audio, starting to measure the signal energy, as the energy in the start of the audio may have hints about pulmonary capacity.
% fiz correçoes gramaticais no parágrafo acima
% mas acho que os  tópicos apresentados (autocorrelaçao?? energia no in[cio sem comparar com outra energia ao longo da fala??) estão um tanto frágeis para já serem considerados passos futuros

\section*{Acknowledgments}
We also gratefully acknowledge the support of NVIDIA corporation with the donation of a GPU used in part of the experiments presented in this research.

\bibliographystyle{apalike}
\bibliography{referencias}

\label{lastpage}
\end{document}